\documentclass[conference]{IEEEtran}
\IEEEoverridecommandlockouts
\usepackage{cite}
\usepackage{amsmath,amssymb,amsfonts}
\usepackage{algorithmic}
\usepackage{graphicx}
\usepackage{textcomp}
\usepackage[table,xcdraw]{xcolor}
\usepackage{enumitem}
\usepackage{pifont}
\usepackage[hyphens]{url}
\usepackage{float}
\usepackage{hyperref}
\newcommand{\cmark}{\ding{51}}%
\newcommand{\xmark}{\ding{55}}%

\def\BibTeX{{\rm B\kern-.05em{\sc i\kern-.025em b}\kern-.08em
    T\kern-.1667em\lower.7ex\hbox{E}\kern-.125emX}}
\begin{document}
\title{Decentralizing Custodial Wallets with MFKDF
}

\author{\IEEEauthorblockN{Vivek Nair
}
\IEEEauthorblockA{\textit{UC Berkeley}\\
Berkeley, CA, USA \\
vcn@berkeley.edu
}
\and
\IEEEauthorblockN{Dawn Song
}
\IEEEauthorblockA{\textit{UC Berkeley}\\
Berkeley, CA, USA \\
dawnsong@berkeley.edu
}
}

\newcommand{\dawn}[1]{Dawn: \textcolor{red}{#1}}

\IEEEoverridecommandlockouts

\IEEEpubid{\makebox[\columnwidth]{978-8-3503-1019-1/23/\$31.00~\copyright2023 IEEE \hfill} \hspace{\columnsep}\makebox[\columnwidth]{ }}

\maketitle

\IEEEpubidadjcol

\begin{abstract}
The average cryptocurrency user today faces a difficult choice between centralized custodial wallets, which are notoriously prone to spontaneous collapse, or cumbersome self-custody solutions, which if not managed properly can cause a total loss of funds. In this paper, we present a ``best of both worlds'' cryptocurrency wallet design that looks like, and inherits the user experience of, a centralized custodial solution, while in fact being entirely decentralized in design and implementation. In our design, private keys are not stored on any device, but are instead derived directly from a user's authentication factors, such as passwords, soft tokens (e.g., Google Authenticator), hard tokens (e.g., YubiKey), or out-of-band authentication (e.g., SMS). Public parameters (salts, one-time pads, etc.) needed to access the wallet can be safely stored in public view, such as on a public blockchain, thereby providing strong availability guarantees. Users can then simply ``log in'' to their decentralized wallet on any device using standard credentials and even recover from lost credentials, thereby providing the usability of a custodial wallet with the trust and security of a decentralized approach.
\end{abstract}

\begin{IEEEkeywords}
mfkdf, custodial wallet, key management, mfa, applied cryptography, cryptocurrency, blockchain, sovereignty
\end{IEEEkeywords}

\section{Introduction}

The recent collapse of several major cryptocurrency exchange platforms offering custodial wallet services \cite{huang_why_2022} has highlighted a critical vulnerability in the decentralized finance ecosystem: an over-reliance on centralized custodians of supposedly decentralized assets. By controlling large portions of cryptocurrency market capitalization \cite{noauthor_coinbase_nodate}, custodial wallets introduce concentrated failure modes into otherwise resilient decentralized platforms \cite{wood_custodial_2022}, thereby unduly reducing public trust in blockchain and crypto technologies as a whole \cite{acheson_after_2022}.

Still, the appeal of custodial wallet services is undeniable. Users, who are notoriously bad at understanding and using public-key cryptography \cite{whitten_usability_nodate}, can simply log in with familiar authentication factors like passwords while enjoying the security advantages of multi-factor authentication (MFA). Should any of these factors be lost, platforms typically provide native support for account recovery using alternative channels like email or SMS. Users benefit from the platforms' fault-tolerant and highly-available architectures, and enjoy the portability of being able to easily access their wallet from any device.

Today, self-custody wallets offer few of these benefits. Users are forced to securely store and manage cryptographic keys, and transact using pseudorandom addresses rather than human-readable identifiers. Their lack of native recovery mechanisms has led to countless notorious cases of millions of dollars of cryptocurrency being lost due to misplaced hardware \cite{noauthor_bitcoin_2022} or forgotten passwords \cite{popper_lost_2021, radio__this_2021}. Cross-device portability is virtually nonexistent, with no way to access funds on a new system without manually transferring keys from an old device.

It is increasingly evident that cryptocurrency users must reduce their reliance on centralized custodians in order to dampen the impact of large failure events and regain public trust in DeFi. However, it is also clear that many are unwilling to move to self-custody solutions until they offer usability on par with custodial wallets. Therefore, the goal of this paper is to provide a trustless, decentralized wallet design that looks and feels like a centralized custodial solution. In doing so, it provides the usability, portability, and recoverability of custodial wallets along with the strong security and privacy advantages of eliminating all trusted parties and committees.

Instead of storing cryptographic keys using hardware or software, our solution utilizes the multi-factor key derivation function (MFKDF) \cite{https://doi.org/10.48550/arxiv.2208.05586} to allow users to derive keys as needed using only familiar authentication factors like passwords, HMAC-based one-time password (HOTP) \cite{rfc4226} and time-based one-time password (TOTP) \cite{rfc6238} codes, out-of-band authentication (OOBA) such as email and SMS, and hardware tokens such as YubiKeys \cite{yubikey}. Users can thus simply ``log in'' to their wallet on any device using a human-readable username and some combination of these factors, as if it was hosted on a centralized exchange. Alternative factor combinations can be established for account recovery in case a primary authentication factor is lost. All public material necessary to facilitate this functionality, such as cryptographic salts and one-time pads, can be stored safely in the open (e.g., on a public blockchain) with no loss in security or privacy.

\subsection*{Contributions}
\noindent To summarize, our proposed wallet design provides several key advantages over existing self-custody solutions:

\begin{enumerate}[leftmargin=*]
    \item We present the first known trustless self-custody wallet design that derives keys from common authentication factors like passwords, HOTP, TOTP, and YubiKeys (\S\ref{sec:mfkdf}).
    \item Our approach provides strong availability guarantees with seamless cross-device portability (\S\ref{sec:availability}).
    \item Users can authenticate and transact using human-readable identifiers rather than unintelligible addresses (\S\ref{sec:usernames}). 
    \item Our system provides built-in support for account recovery using secondary factors like email and SMS (\S\ref{sec:recovery}).
\end{enumerate}

\section{Background \& Related Work}

Between hardware, software, and custodial solutions, there are hundreds of cryptocurrency wallets for users to choose from today \cite{corva_13_nodate}. The purpose of this section is to outline the current state of cryptocurrency wallet design according to literature reviews of the field \cite{9315193, eyal2022cryptocurrency}, so as to clearly differentiate the proposed approach from known techniques.

\subsection{Custodial Wallets}

Custodial cryptocurrency wallets, whereby a centralized third-party service provider is trusted to store and manage private keys on behalf of their users, remain one of the most popular ways to enter the cryptocurrency space. In addition to the aforementioned advantages of usability, portability, and recoverability, custodial wallets are usually offered in conjunction with a centralized exchange where fiat currency can be used to purchase cryptocurrency assets. 

The three largest centralized platforms, Binance \cite{binance}, Coinbase \cite{coinbase}, and Kraken \cite{kraken}, together account for nearly \$20 billion in daily trading volume \cite{top_cex}, orders of magnitude larger than the largest decentralized exchanges \cite{dex}.

Unfortunately, this centralization has also led custodial wallets to be overrepresented in their share of major security incidents and fraud.
From the infamous collapse of Mt. Gox in 2014 \cite{mt_gox} to the recent downfall of FTX \cite{ftx}, custodial services have proven notoriously prone to catastrophic failure. Using such services fundamentally alters the trust assumptions underlying decentralized systems, defeating many of their security and privacy benefits. We are thus motivated to explore self-custody solutions, which are the focus of this paper.

\subsection{Hardware Wallets}

Hardware wallets, such as those offered by Ledger \cite{ledger} and Trezor \cite{trezor}, are considered the gold standard for the safe management of cryptocurrency due to their use of a secure element for storing private keys and the absence of a general computing device which may be susceptible to malware.

Unfortunately, these devices also lie at the opposite end of the usability spectrum due to their relatively high up-front cost, cumbersome physical interface, and intrinsic limitations on the number and types of cryptocurrencies which can be used \cite{9315193}.

Hardware wallets offer a complete lack of portability, with no easy way to access funds on any device that is not physically connected to the wallet.
The absence of built-in redundancy means that the loss of a single device can result in a total loss of funds unless additional work is done to establish backups. Accordingly, the community regularly hears stories of millions of dollars of cryptocurrency being lost due to misplaced hardware \cite{noauthor_bitcoin_2022, browne_man_nodate}.
Furthermore, most hardware wallets offer no native recovery mechanism other than the use of a BIP39 seed phrase \cite{bip39}, which must itself be stored and managed securely and is equally susceptible to loss.

Given these deficiencies, hardware wallets present an onerously high barrier to entry for most cryptocurrency users. We next turn to software-based solutions which aim to address these shortcomings but carry their own set of drawbacks.

\subsection{Software Wallets}

Non-custodial software-based wallets are a popular alternative to dedicated hardware wallets due to their lower entry cost and improved usability.
While some literature reviews consider mobile wallets, such as Trust Wallet \cite{trustwallet}, and desktop wallets, such as MetaMask \cite{metamask}, to be entirely separate categories \cite{9315193}, software wallets operate using fundamentally similar mechanisms regardless of platform.

The conventional software wallet design involves storing a private key file directly on a user's file system, usually encrypted using a password. Portability is achieved by allowing this key file to be moved from one machine to another. Thus, the security of this setup is reduced to the ability to choose a secure password and manage private keys securely, both difficult tasks for the average user \cite{whitten_usability_nodate, password_reuse, florencio_large_2006, credential_stuffing}.

Like with hardware wallets, a BIP39 seed phrase is often the only supported recovery mechanism, and cases of lost funds due to forgotten passwords are widespread \cite{popper_lost_2021, radio__this_2021}.

\subsection{Multi-Factor Authentication}
Within centralized applications, multi-factor authentication (MFA) has long been the go-to solution for both the insecurity of passwords as a sole authentication factor and the problem of account recovery. 
While a variety of MFA mechanisms are currently in use, one-time passwords (OTPs), such as HOTP \cite{rfc4226}, TOTP \cite{rfc6238}, and OOBA \cite{ooba}, are amongst the most popular MFA methods in use today.

Centralized exchanges like Coinbase \cite{coinbase} support and encourage the use of all of these methods, yet the direct protection of non-custodial cryptocurrency wallets using OTP factors is rarely seen in practice. Doing so in a cryptographically-secure way would require private keys to be locally derived from all authentication factors instead of just passwords, a feat which has not, until recently, been thought possible \cite{https://doi.org/10.48550/arxiv.2208.05586}. 

Unlike passwords, which are expected to remain fairly constant over time, OTPs are, by definition, intended for one-time use, and are thus expected to change upon each login. It is not immediately clear how a key can be deterministically derived from the OTP corresponding to any given login request. Instead, current attempts to construct MFA-based wallets rely either on new, purpose-built authentication factors, or on secret sharing keys across semi-trusted committees.

\subsection{MFA Wallets}
The idea of using multi-factor authentication to secure a cryptocurrency wallet is not entirely new. In fact, several works have proposed new MFA protocols specifically for the purpose of securing cryptocurrency wallets \cite{8644084, homoliak2018air}. However, the usability advantages of deriving private keys from familiar factors are somewhat diminished if users are required to learn and adopt a new MFA method in order to use a wallet.

Alternatively, several cryptocurrency wallet solutions have been proposed which rely upon secret-sharing a private key across a committee of nodes, at least some threshold of which are presumed to be honest \cite{8271996, zhu2017trust}. Users can then authenticate with these nodes using standard authentication factors to recover the shares of their key, providing a semi-custodial experience.
While more resilient than relying on a single trusted entity, these solutions do not provide the security properties of a fully-decentralized approach, in part because the nodes constituting a trusted committee are often homogeneous in design and thus subject to common vulnerabilities. By contrast, the wallet design of this paper relies neither on trusted third parties nor on partially honest committees, while also potentially allowing users to choose their own trust models through support for flexible policies (see \S\ref{sec:future}).

Deriving cryptocurrency wallet keys from popular, unmodified authentication factors like HOTP and TOTP without relying on a trusted third party or committee has long been considered a hard problem due to the dynamic nature of OTPs not being readily conducive to the derivation of a static key. However, MFKDF is designed to solve this exact problem, and by leveraging this technique, we are the first to present a cryptocurrency wallet design that is both completely trustless and fully backward-compatible with popular MFA factors.

\subsection{MFKDF}
The Multi-Factor Key Derivation Function (MFKDF) \cite{https://doi.org/10.48550/arxiv.2208.05586} is a recent improvement over password-based key derivation that incorporates multiple authentication factors into the key derivation process. Its construction provides the fundamental building block for the creation of a decentralized multi-factor authenticated wallet with support for standard, unmodified authentication factors such as HOTP and TOTP. Through a unique ``key-feedback mechanism,'' MFKDF allows for the secure derivation of a static key from dynamic OTP factors.

MFKDF takes as input dynamic \textit{factor witnesses} ($W$) and public parameters ($\alpha$) and produces static key material ($\sigma$). 
\textit{Factor witnesses} ($W$) refer to the exact values provided by a user to authenticate (e.g., a password and a 6-digit OTP). The public parameters ($\alpha$), which include values like cryptographic salts and one-time pads, require no security assumptions and can safely be stored in the public, such as on a public blockchain. In most cases, these parameters must be updated upon each login ($\alpha_i \mapsto \alpha_{i+1}$). Therefore, a significant focus of this paper will be on storing and updating the public parameters ($\alpha$) in a secure, portable, resilient, and highly-available manner. The resulting output ($\sigma$) can then be used to derive one or more cryptocurrency wallet private keys.

In addition to providing secure key derivation from multiple authentication factors, MFKDF also elegantly addresses the key recovery problem through a threshold mechanism that allows $n$ authentication factors to be established, only some of which ($t < n$) are actually required to derive the key. As such, some number of authentication factors can be lost without causing the key to be lost entirely. In doing so, it allows primary authentication factors (such as a password and TOTP code) to be used for normal key derivation, and secondary authentication factors (such as email or SMS) to be used to recover the key when a primary factor, providing a very similar user experience to centralized applications.

\subsection{Summary}
In conclusion, cryptocurrency users today are still stuck with a difficult choice between convenient but fallible custodial wallets and secure but cumbersome self-custody solutions. MFKDF allows us, for the first time, to provide a ``best of both worlds'' non-custodial solution with the interface and user experience of a centralized service.
In the following section, we more precisely describe the security and privacy goals of our solution before describing the specification in \S\ref{sec:system}.

\vspace{0.5em}
\begin{table}[H]
\resizebox{\columnwidth}{!}{%
\begin{tabular}{l|c|c|c|}
\cline{2-4}
 & {\color[HTML]{333333} \textbf{\begin{tabular}[c]{@{}c@{}}Custodial\\ Wallets\end{tabular}}} & {\color[HTML]{333333} \textbf{\begin{tabular}[c]{@{}c@{}}Non-Custodial\\ Wallets\end{tabular}}} & {\color[HTML]{333333} \textbf{\begin{tabular}[c]{@{}c@{}}MFKDF\\ Wallet\end{tabular}}} \\ \hline
\multicolumn{1}{|l|}{{\color[HTML]{333333} Decentralized}} & {\color[HTML]{9B9B9B} \xmark} & {\color[HTML]{009901} \cmark} & {\color[HTML]{009901} \cmark} \\ \hline
\multicolumn{1}{|l|}{{\color[HTML]{333333} Trustless}} & {\color[HTML]{9B9B9B} \xmark} & {\color[HTML]{009901} \cmark} & {\color[HTML]{009901} \cmark} \\ \hline
\multicolumn{1}{|l|}{{\color[HTML]{333333} Portable}} & {\color[HTML]{009901} \cmark} & {\color[HTML]{9B9B9B} \xmark} & {\color[HTML]{009901} \cmark} \\ \hline
\multicolumn{1}{|l|}{{\color[HTML]{333333} Resilient}} & {\color[HTML]{009901} \cmark} & {\color[HTML]{9B9B9B} \xmark} & {\color[HTML]{009901} \cmark} \\ \hline
\multicolumn{1}{|l|}{{\color[HTML]{333333} MFA}} & {\color[HTML]{009901} \cmark} & {\color[HTML]{9B9B9B} \xmark} & {\color[HTML]{009901} \cmark} \\ \hline
\multicolumn{1}{|l|}{{\color[HTML]{333333} Familiar Factors}} & {\color[HTML]{009901} \cmark} & {\color[HTML]{9B9B9B} \xmark} & {\color[HTML]{009901} \cmark} \\ \hline
\multicolumn{1}{|l|}{{\color[HTML]{333333} Recoverable}} & {\color[HTML]{009901} \cmark} & {\color[HTML]{9B9B9B} \xmark} & {\color[HTML]{009901} \cmark} \\ \hline
\end{tabular}%
}
\vspace{1em}
\caption{Properties of custodial, non-custodial, and MFKDF wallets.}
\label{tab:comparison}
\end{table}
\vspace{-1.5em}

\section{Problem Statement}
\label{sec:problem}

For simplicity, this paper assumes that a cryptocurrency wallet is defined by a single private key ($\sigma$) and belongs to a single authorized user. That user is the only party in posession of the authentication factors used to constitute their wallet, and is thus the only party able to efficiently produce \emph{factor witnesses} ($W$) for those factors. No assumptions are made of the entities storing public parameters ($\alpha$) for a wallet. The basic security requirement of our system is that only the authorized user is able to obtain the private key ($\sigma$) of the wallet. Further desired properties of our system are as follows:

\begin{enumerate}[leftmargin=*]
    \item \emph{Decentralized}: No external third party or committee shall have custodial access to a user's private wallet keys.
    \item \emph{Trustless}: The correct operation of the wallet shall not rely on the honest behavior of any third party or committee.\footnote{In some cases, malicious colluding parties could perform an eclipse attack, leading to an unavoidable temporary denial of service for a legitimate user.}
    \item \emph{Portable}: Users can access their wallet on new devices regardless of the availability of previously-used devices.
    \item \emph{Resilient}: The availability a wallet does not depend on accessing a small number of specific physical devices.
    \item \emph{Multi-Factor Authenticated}: Wallet security (entropy) is a direct product of all authentication factors used.
    \item \emph{Compatible}: Familiar, unmodified authentication factors (e.g., TOTP, OOBA) can be used to access the wallet.
    \item \emph{Recoverable}: Alternative factors (e.g., SMS, email) can be used to recover the wallet with equivalent availability.
\end{enumerate}

As shown in Table \ref{tab:comparison}, most cryptocurrency wallets in existence can offer only a subset of the desired properties. Existing attempts at building multi-factor authenticated wallets are either committee-based (not trustless) or require new, purpose-built authentication methods (not compatible with familiar factors). In the next section, we will describe an MFKDF-based wallet architecture that meets the stated goals.
\section{System Overview}
\label{sec:system}

\subsection{Multi-Factor Key Derivation}
\label{sec:mfkdf}

\begin{figure}[H]
\includegraphics[width=0.75 \linewidth]{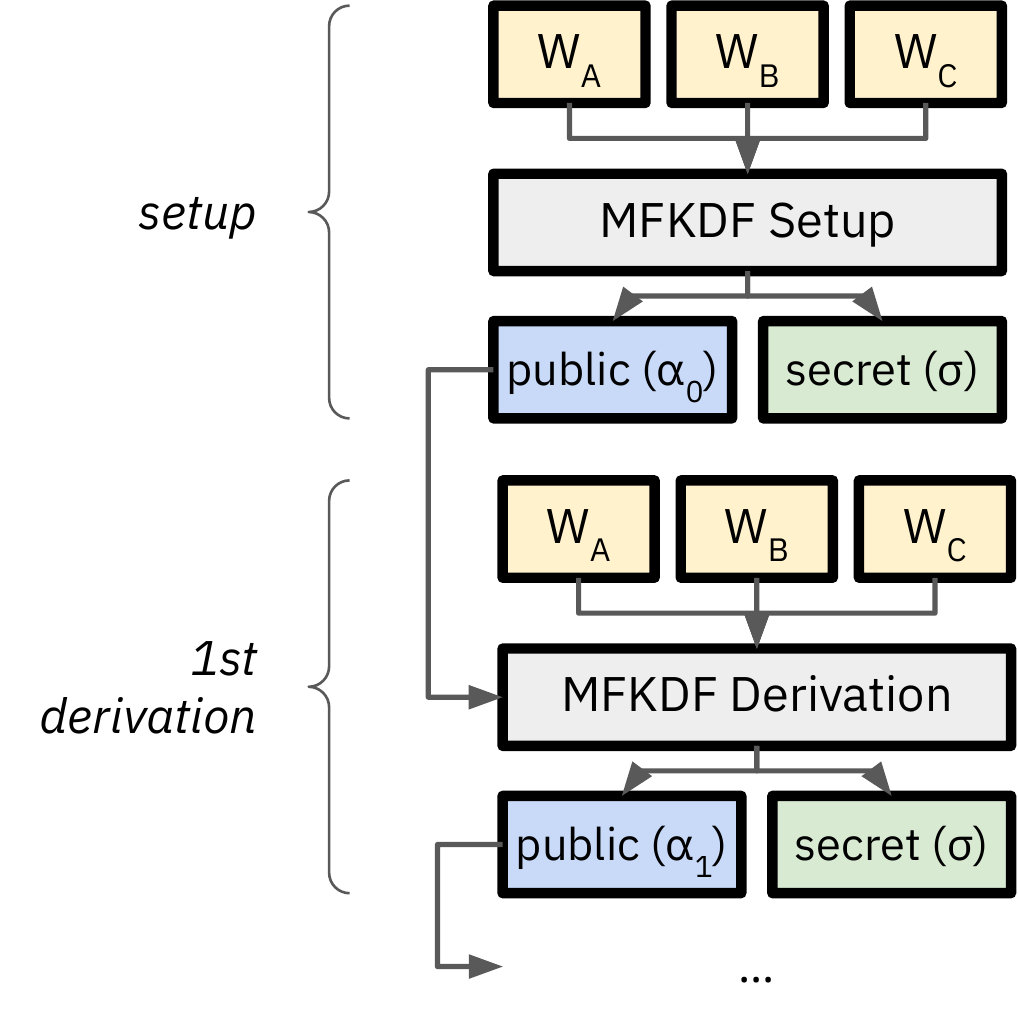}
\centering
\caption{Basic usage pattern of MFKDF setup and derive functions.}
\label{fig:system_mfkdf}
\end{figure}

The fundamental technique of this paper is to use the multi-factor key derivation function (MFKDF) \cite{https://doi.org/10.48550/arxiv.2208.05586} to derive wallet keys dynamically from standard authentication factors rather than storing private keys in any location. The MFKDF specification (simplified here for comprehensibility) is split into setup and derive functions, as shown in Fig. \ref{fig:system_mfkdf}. The setup function takes as input several authentication factors ($W$) and produces public parameters ($\alpha$) and a secret key ($\sigma$). The public parameters must be used in the subsequent derive function along with the same factors in order to produce the same secret key, and will also be updated in the process.

This parameter feed-forward mechanism of MFKDF is an important innovation that allows standard authentication factors like HOTP, TOTP, YubiKey, and OOBA to be used as part of the key derivation process. As such, a user can ``sign up'' using standard authentication factors, producing a private wallet key and public parameters, which can be safely stored. Later, those public parameters can be used along with the same authentication factors to ``log in'' to the wallet, causing the same private key to be derived if and only if correct authentication factors are provided. Thus, the key management aspect of the wallet is obscured, with the user experience matching that of a custodial service. Simultaneously, the user enjoys the security advantages of MFA, with the entropy of the key being jointly derived from all used authentication factors.

\subsection{Availability \& Portability}
\label{sec:availability}

\begin{figure}[H]
\includegraphics[width=0.75 \linewidth]{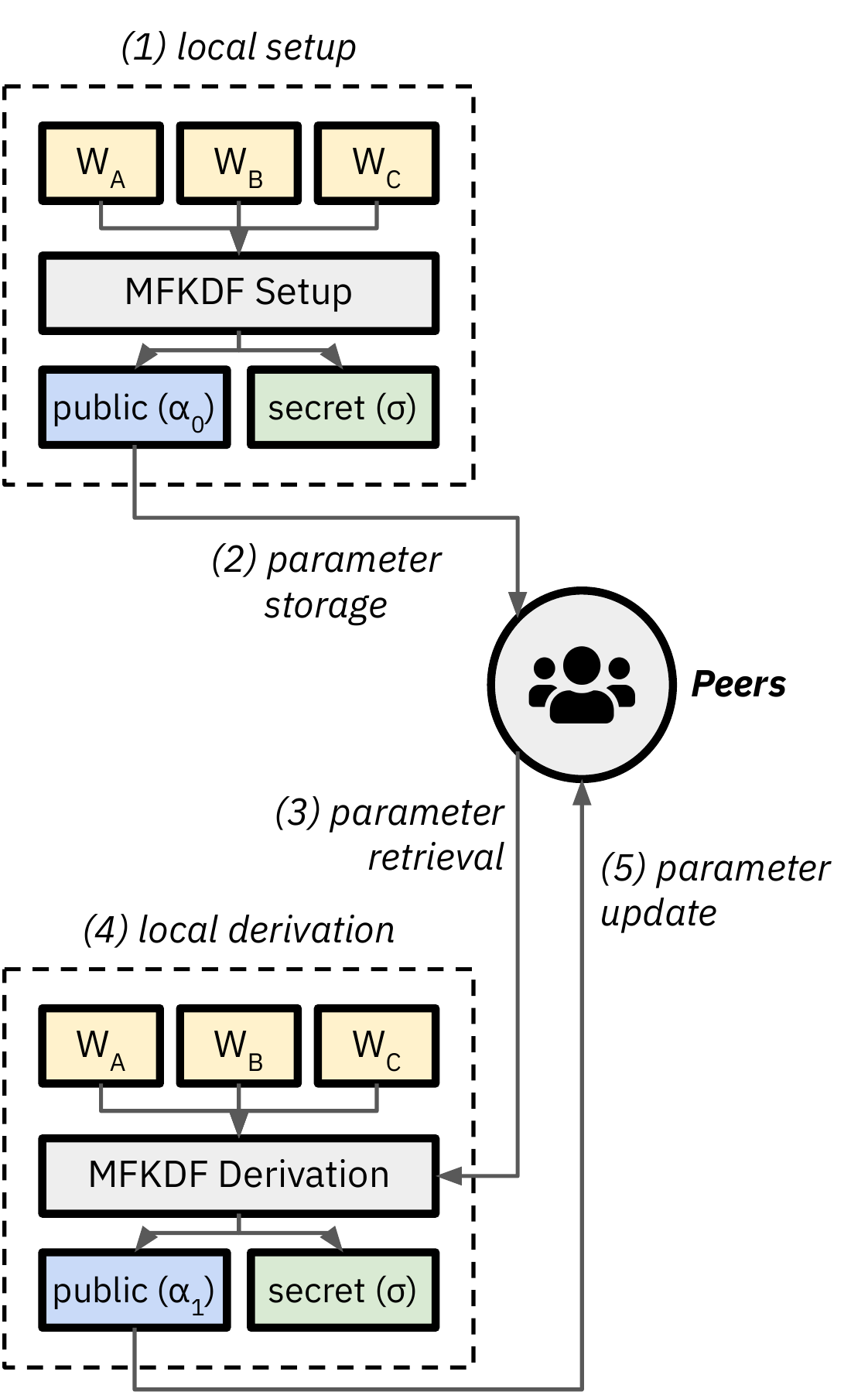}
\centering
\caption{Distributed system for storing public MFKDF values.}
\label{fig:system_portability}
\end{figure}

Although we have thus far succeeded at producing a self-custody cryptocurrency wallet that is secured by multi-factor authentication, the use of locally-stored public parameters in the derivation process limits the portability of this approach. In this section, we focus on ensuring the public parameters are always available for login on any device when desired.
Because no trust assumptions are required of the party storing public material, even a public blockchain could be used to store these values. However, due to the prohibitively high cost of doing so, we instead suggest the use of a simple peer-to-peer storage approach, as shown in Fig. \ref{fig:system_portability}.

Per the suggested approach, the wallet application forms a network with its peers, and stores a copy of all legitimate MFKDF public parameters. During a local setup process, (1) an MFKDF key is established and (2) the public parameters are broadcast to the network. Later, when the wallet is accessed on another device, (3) the parameters are retrieved from the network and (4) combined with the authentication factors to re-derive the wallet key. Finally, (5) the parameters are updated and the new parameters are stored on the network. Several existing networks, such as IPFS \cite{benet2014ipfs}, can serve this purpose.

\begin{figure}[H]
\includegraphics[width=0.6 \linewidth]{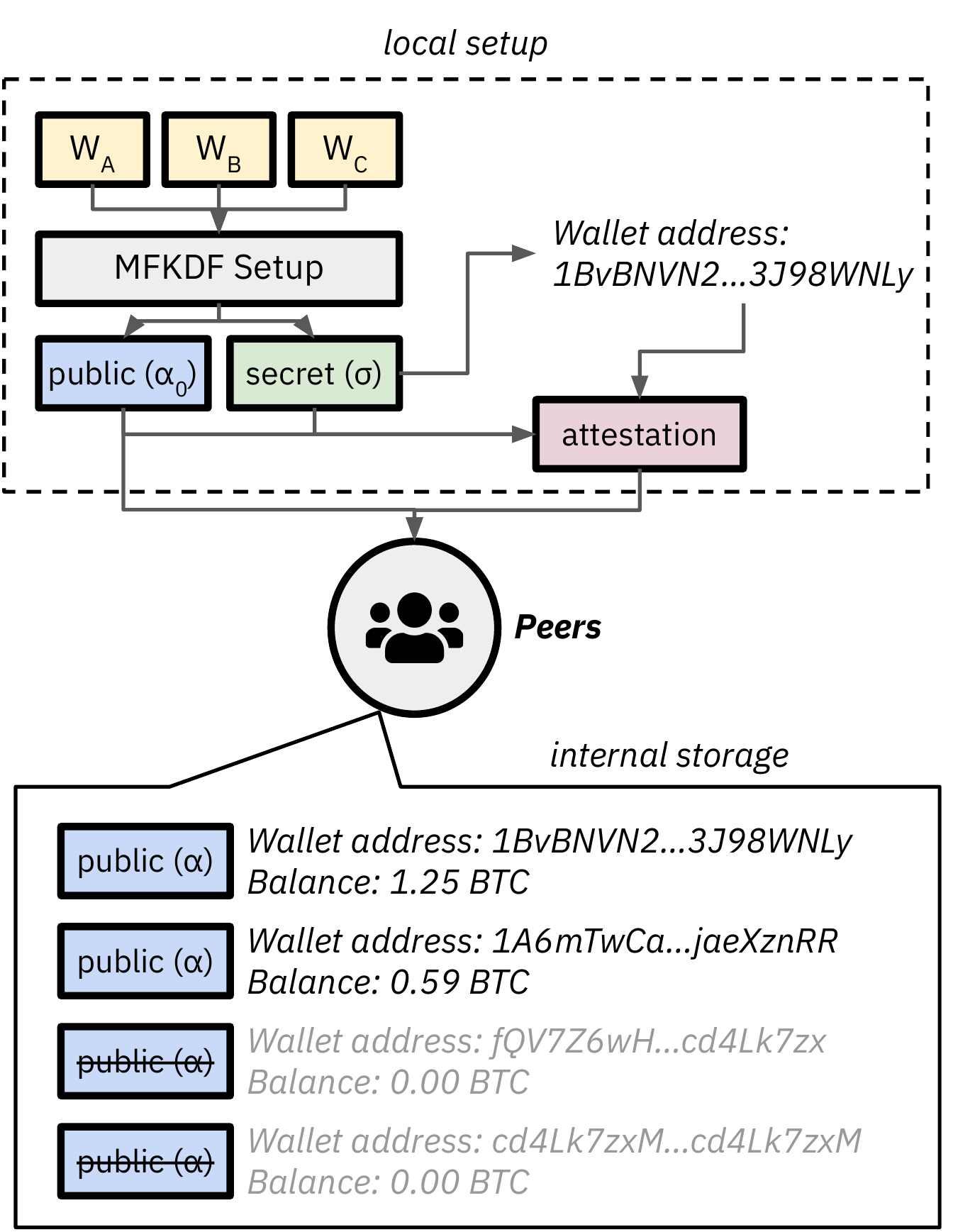}
\centering
\caption{Attestation mechanism for linking wallet addresses to public material.}
\label{fig:system_attestation}
\end{figure}

The relatively small size ($\leq 10$~kb) of MFKDF public parameters corresponding to each wallet makes this approach concretely practical, and it can be improved further through techniques like sharding as discussed in \S\ref{sec:future}. However, a significant drawback of the method as it stands is the potential ability for adversaries to flood the network with worthless (empty) wallets, thus exhausting available storage space and causing a denial of service for legitimate users. Therefore, we also suggest the use of an attestation mechanism that links wallet addresses to stored public material, as shown in Fig. \ref{fig:system_attestation}.

Per the example of Fig. \ref{fig:system_attestation}, MFKDF public material documents are signed using cryptocurrency wallet keys, allowing participants to identify the wallet associated with each stored public material object as well as its on-chain balance. If space is limited, material corresponding to addresses with zero on-chain asset value can be discarded without destruction of value, thus preventing denial-of-service attacks while ensuring high availability for legitimate (valuable) wallets.

\subsection{Human-Readable Identifiers}
\label{sec:usernames}

While the above system and network architecture succeeds at offering high availability and preventing denial of service, it still requires users to remember long, pseudorandom wallet addresses in order to retrieve their public material and access their wallet. By contrast, a major usability advantage of centralized platforms is the ability to access accounts using a human-readable identifier, usually an email address. Thankfully, email authentication is a form of OOBA natively supported by MFKDF \cite{https://doi.org/10.48550/arxiv.2208.05586}, thereby allowing for email addresses to be used to securely index stored public parameters.

\begin{figure}[H]
\includegraphics[width=0.6 \linewidth]{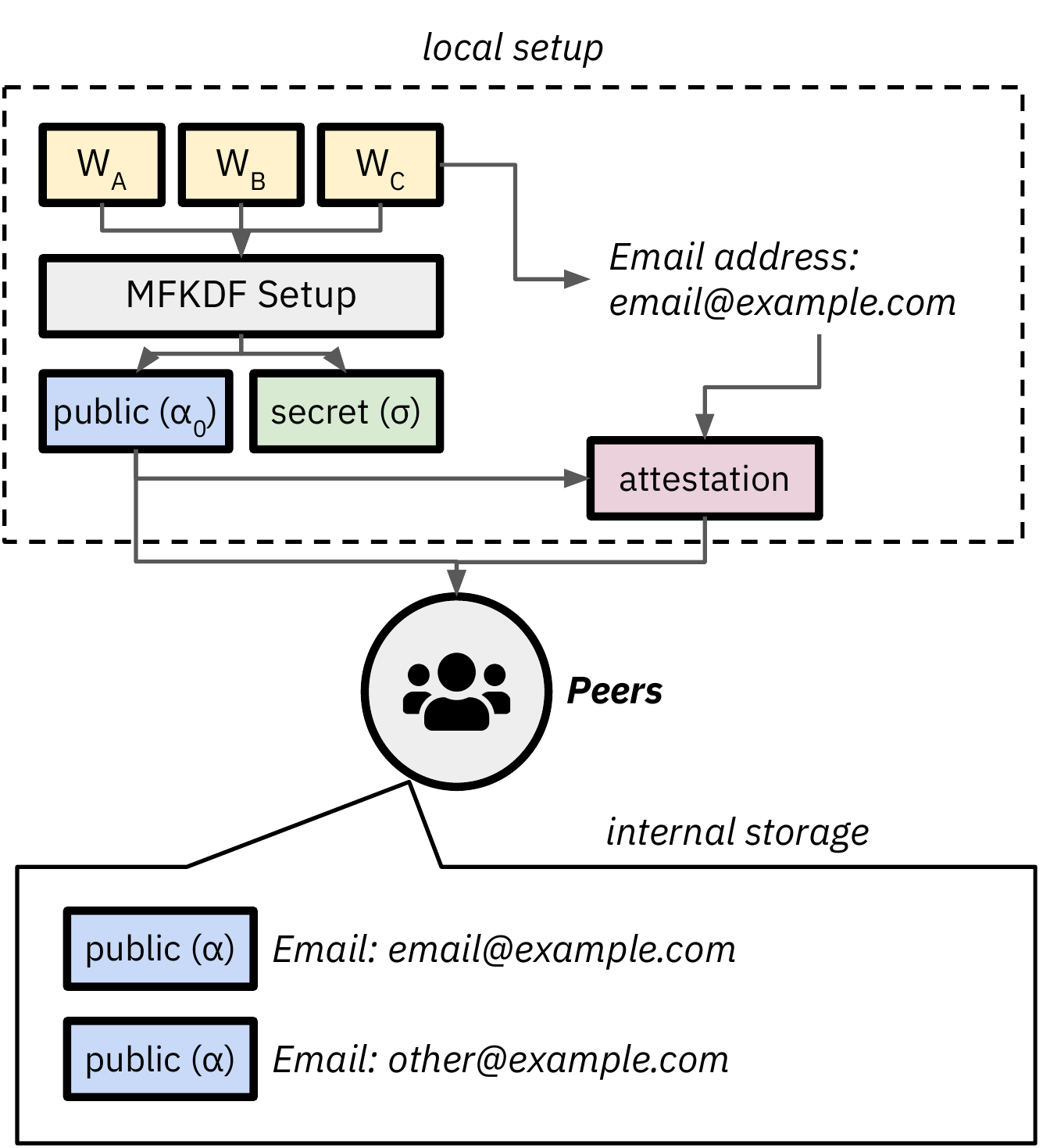}
\centering
\caption{Use of email OOBA for human-readable identifiers.}
\label{fig:system_email}
\end{figure}

As illustrated in Fig. \ref{fig:system_email}, email OOBA is established as one of the factors constituting the MFKDF-derived wallet key. At the end of the setup process, the email factor material is used to produce an attestation of the stored public material, allowing participants to securely store and retrieve email addresses and public material as key-value peers. Users can then ``log in'' to their wallet using an email address, rather than a wallet address, along with their password and other authentication factors, faithfully replicating the centralized wallet experience.

\subsection{Account Recovery}
\label{sec:recovery}
Thus far, our proposal has centered around the simplest form of multi-factor key derivation, whereby $n$ separate authentication factors are established, and all $n$ are required to later derive the key. However, the MFKDF specification also provides for a threshold variant of MFKDF \cite{https://doi.org/10.48550/arxiv.2208.05586}, whereby $n$ factors are established, only $t$ of which are later required to derive the key ($0 < t < n$). Behind the scenes, Shamir's secret sharing \cite{sss} is used to ensure that the key can be derived if and only if any $t$ of the established factors are correctly provided.

\begin{figure}[H]
\includegraphics[width=0.7 \linewidth]{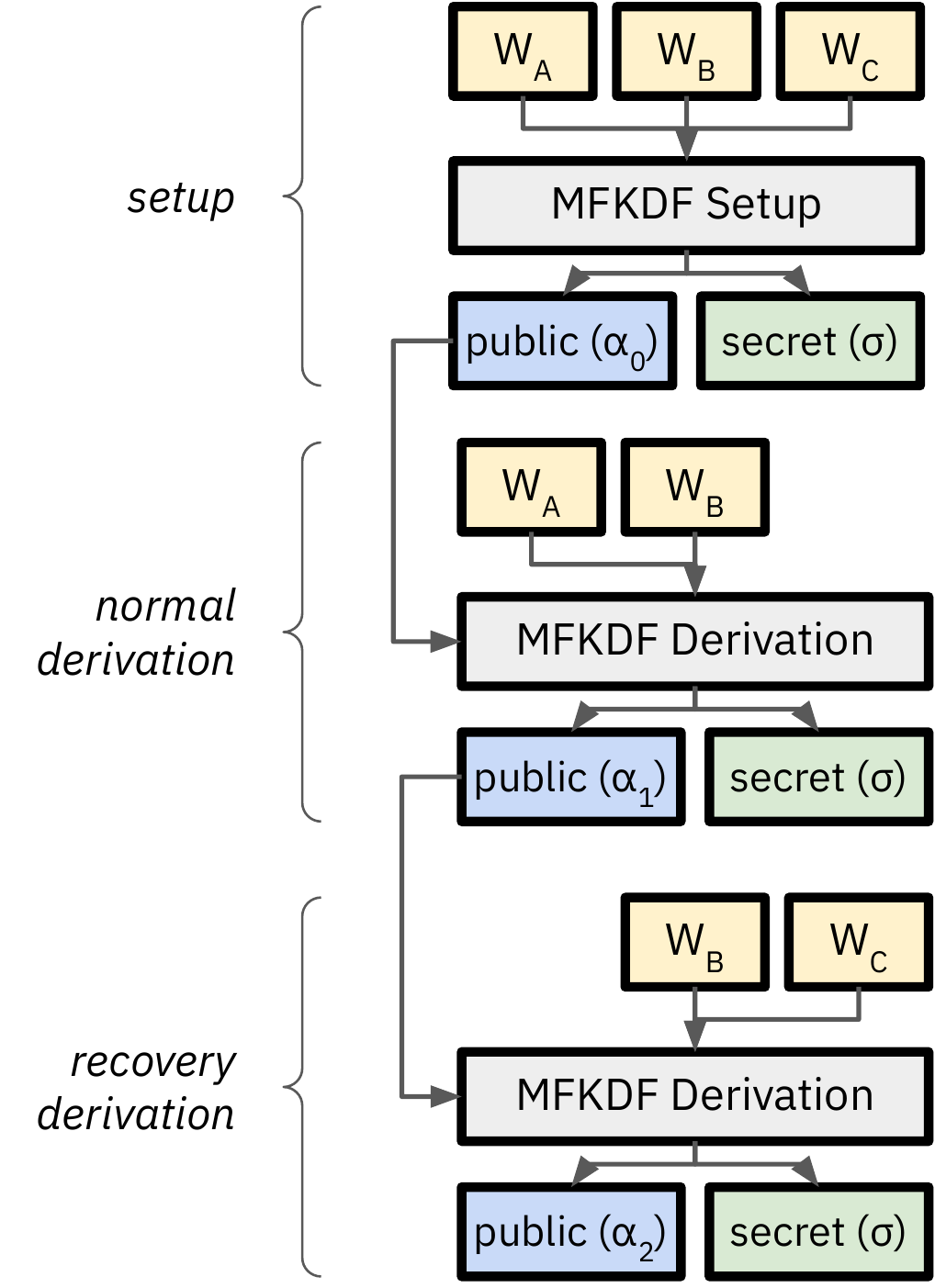}
\centering
\caption{Account recovery using threshold MFKDF.}
\label{fig:system_recovery}
\end{figure}

The threshold version of MFKDF can be used to seamlessly facilitate account recovery in the event of a lost factor, as shown in Fig. \ref{fig:system_recovery}. For instance, password, TOTP, and email OOBA factors can be used to establish an MFKDF-derived key according to a $2$-of-$3$ threshold setup. During normal wallet accesses, a password and TOTP code can be supplied to derive the wallet key. However, if either the password or TOTP code is lost, the email OOBA factor can be used together with the remaining factor to recover the wallet. This setup is completely analogous to the typical account recovery process in centralized platforms, providing a familiar user experience and resilience to lost factors without weakening security.
\clearpage

\section{Proof-of-Concept Implementation}

To demonstrate the immediate practical utility of our MFKDF-based wallet architecture and provide a blueprint for its deployment, we implemented a functional MFKDF-based Ethereum web wallet application using the existing JavaScript library for MFKDF \cite{mfkdf} and the eth-hot-wallet project \cite{eth-hot-wallet}. The implementation uses a React.js front-end and operates as a light wallet, and must be connected to a full node to send and receive transactions. As such, no web back-end is required whatsoever, and the entire application can be delivered serverlessly, such as via IPFS.

\subsection{Authentication}
\label{sec:wallet_authentication}

As described in \S\ref{sec:mfkdf}, the main advantage of using MFKDF in a decentralized wallet implementation is that users can ``log in'' to their wallet with traditional authentication factors as if it were an account on a centralized custodial service, without relying on committees or trust assumptions. For our implementation, we chose to use a $2$-of-$3$ threshold MFKDF setup based on a password, YubiKey, and UUIDv4 recovery code, though any of the other authentication factors supported by MFKDF (e.g., HOTP, TOTP, OOBA) could have just as easily been used in place of the chosen factors.

\begin{figure}[H]
\includegraphics[width=0.6 \linewidth]{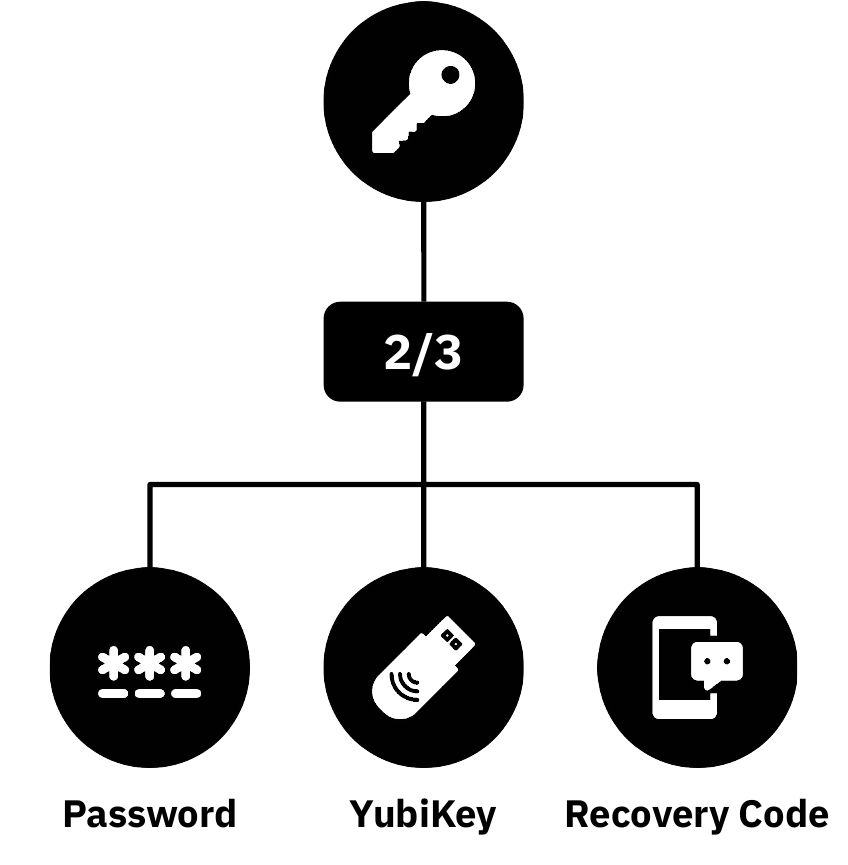}
\centering
\caption{Threshold MFKDF setup used in proof-of-concept implementation}
\label{fig:authentication}
\end{figure}

As shown in Fig. \ref{fig:authentication}, the $2$-of-$3$ threshold key derivation approach allows any of the three factors to be forgotten without losing access to the underlying wallet key. We illustrate how this recovery process works in \S\ref{sec:wallet_recovery}. Given that the average password provides $40$ bits of entropy \cite{florencio_large_2006}, UUIDv4 recovery codes provide 122 bits, and YubiKey via HMAC-SHA1 provides 160 bits, this key derivation policy provides at least 162 bits of security in the weakest configuration. In combination with the chosen KDF (Argon2 \cite{argon2}), the wallet should provide very robust resistance to brute-force attacks.

Upon creating an ``account'' for the wallet, a password is selected, and a recovery code and HMAC secret for YubiKey are randomly generated. The MFKDF setup function is invoked to produce public parameters and a derived key, which is then used to display an Ethereum wallet address. The same wallet can be accessed using the parameters, password, and YubiKey.

\subsection{Functionality}

\begin{figure}[H]
\includegraphics[width=\linewidth]{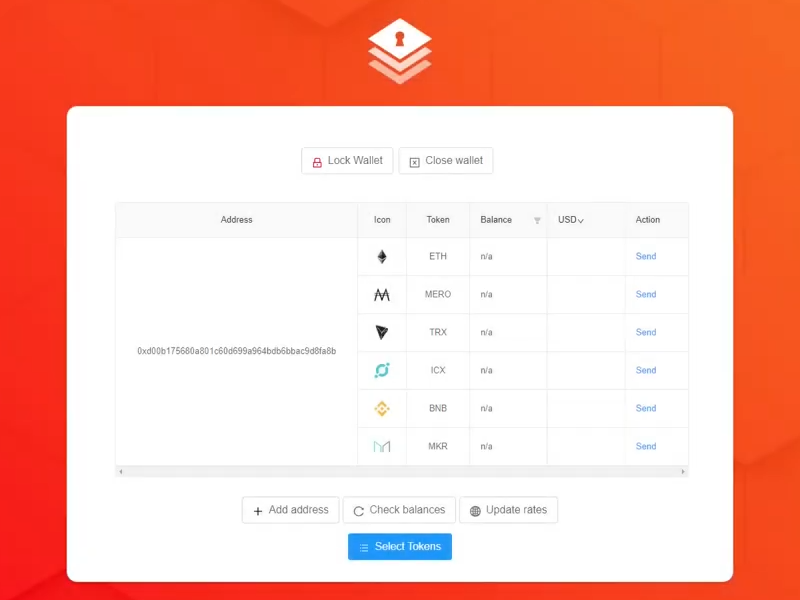}
\centering
\caption{User interface of proof-of-concept wallet implementation}
\label{fig:functionality}
\end{figure}

Fig. \ref{fig:functionality} shows the user interface of our proof-of-concept application after the user has successfully ``signed in'' (derived a key). Using their MFKDF-derived wallet key, the user is able to send and receive Ether as well as a number of ERC-20 tokens. We verified that this functionality was working correctly on both the Ropsten test network and the Ethereum main network. We do not expect difficulty adding other cryptocurrencies like Bitcoin to this wallet in the future.

\vspace{1em}

\subsection{Networking}
\begin{figure}[H]
\includegraphics[width=\linewidth]{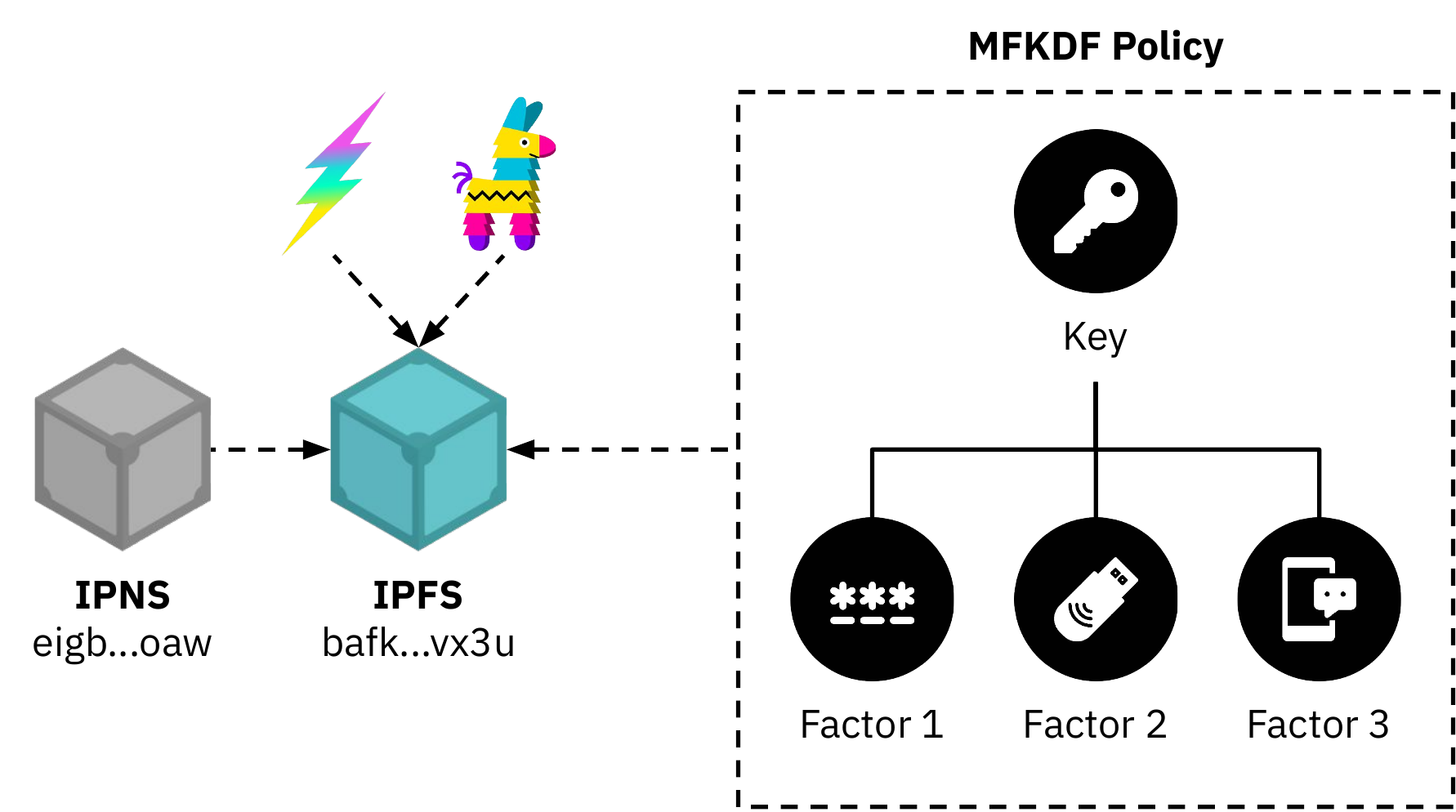}
\centering
\caption{Network architecture of proof-of-concept wallet implementation}
\label{fig:networking}
\end{figure}

Fig. \ref{fig:networking} illustrates the network and distributed system setup used in our proof-of-concept demo. Due to the current lack of a persistent user base, we bootstrapped our system using the InterPlanetary File System (IPFS) \cite{benet2014ipfs} and InterPlanetary Name System (IPNS) \cite{fotiou2021enabling} rather than using the peer-to-peer approach of \S\ref{sec:availability} and the human-readable usernames of \S\ref{sec:usernames}. While slightly less usable than the proposed approach, this was a necessary short-term concession to avoid requiring other MFKDF wallet users to be online for availability.

Upon creating a new wallet, the public material is uploaded to IPFS, and a corresponding IPNS record is created, the address of which becomes the ``username.'' Because other MFKDF wallet users are not currently available to store the public material, we invoke IPNS pinning services such as Pinata \cite{pinata} and Fleek \cite{fleek} to ensure the persistence of the public material. While such services do reintroduce an element of centralization in the short term, they do not require additional trust and only exist to ensure high availability.

\begin{figure}[H]
\includegraphics[width=\linewidth]{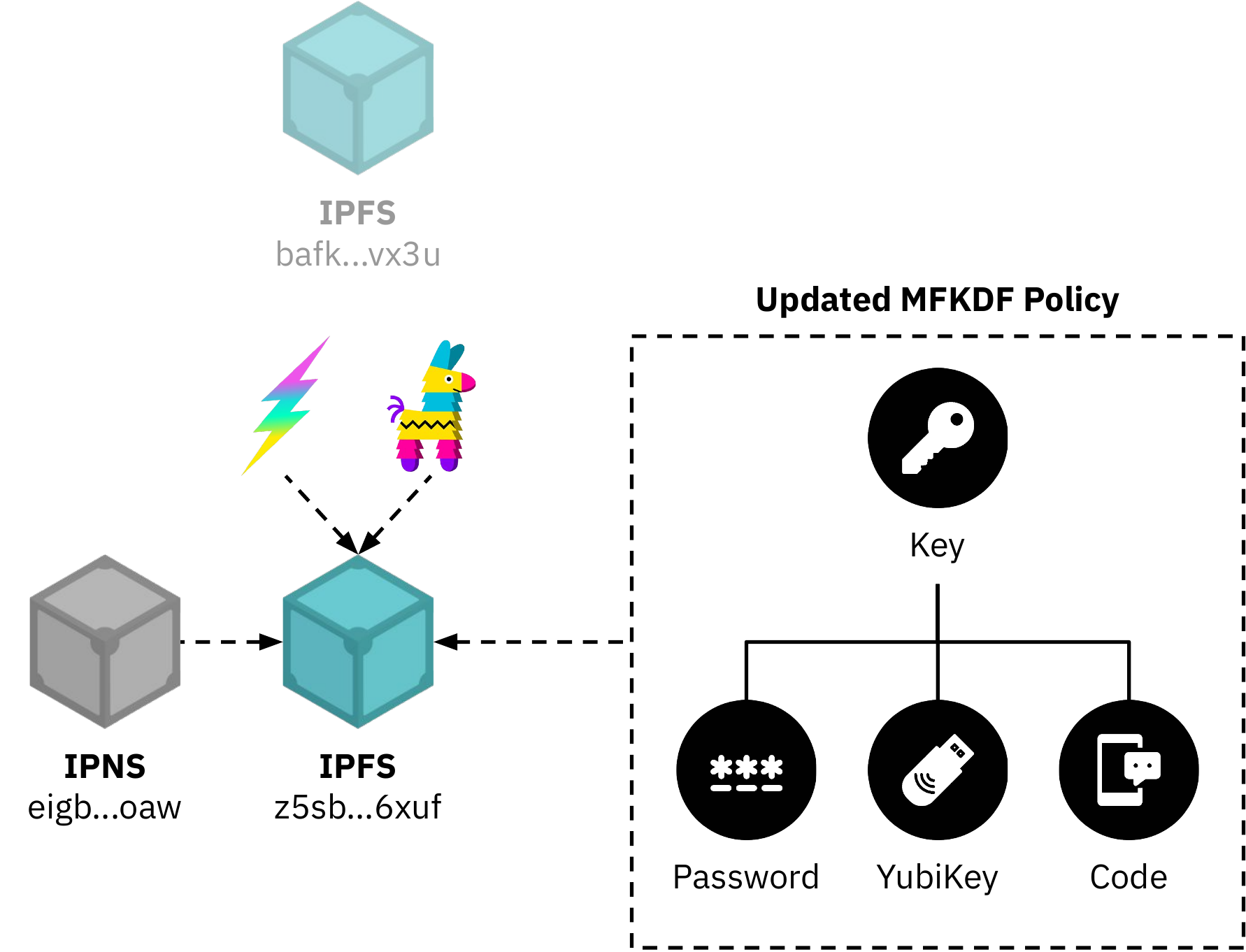}
\centering
\caption{Network updates upon change to MFKDF public parameters}
\label{fig:change}
\end{figure}

When ``logging in'' to their wallet, a user provides their ``username'' (IPNS address, for this demo), along with at least two of their three authentication factors, such as their password and YubiKey. The MFKDF policy document is then updated, and the new policy is uploaded to IPFS. The IPNS record and pins will be updated accordingly, as shown in Fig. \ref{fig:change}, such that the latest parameters are always used to access the wallet. The outdated version of the parameters will then go abandoned, and will stop being stored by other parties shortly thereafter.

\begin{figure}[H]
\includegraphics[width=\linewidth]{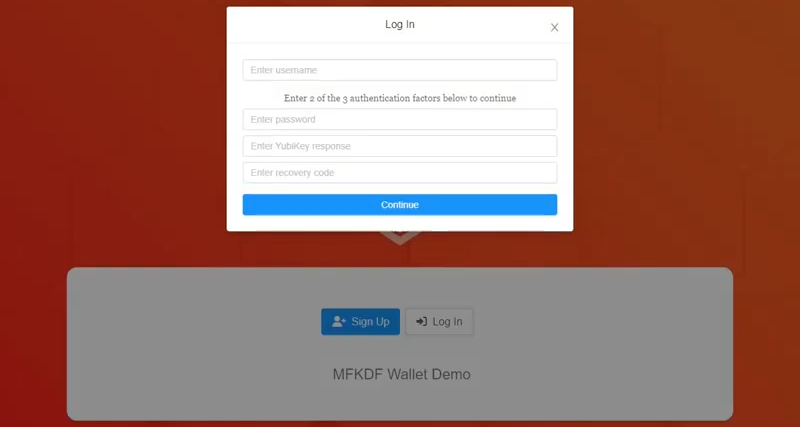}
\centering
\caption{Authentication interface of proof-of-concept wallet implementation}
\label{fig:recovery}
\end{figure}

\subsection{Recovery}
\label{sec:wallet_recovery}
Fig. \ref{fig:recovery} shows the login interface of the wallet demo application. As described in \S\ref{sec:wallet_authentication}, a $2$-of-$3$ threshold MFKDF setup was used to facilitate account recovery in case of a lost factor. For simplicity, we implemented a streamlined interface for login and recovery, whereby any two factors can be used to authenticate at any time. However, for a more traditional user experience, a login page can be configured to request only primary authentication factors, with a separate recovery flow used in case a primary factor is lost. 

\subsection{Discussion}
We present the fully-featured web wallet application demo of this section to illustrate that the proposed MFKDF-based wallet solution is concretely practical and suitable for real-world deployment. While some concessions were made due to the current lack of a persistent install base, the application already largely looks and feels like a centralized custodial wallet, with ``sign up'' and ``log in'' options rather than needing manual key management, while in fact being completely decentralized in design and implementation.

We further demonstrated the backward compatibility of our solution with existing popular MFA methods such as YubiKey, with other methods supported by MFKDF like HOTP, TOTP, and OOBA easily being able to take its place. The chosen factor combination provides excellent brute-force resistance with at least 162 bits of entropy on average, while facilitating factor recovery using a $2$-of-$3$ threshold approach.

Overall, while already practical and advantageous over existing wallets, the usability of our solution is likely to only improve over time with increased adoption and further technological advancements. Features like human-readable identifiers (\S\ref{sec:usernames}) that rely on a degree of existing adoption would serve to only further improve the experience of wallet users.

Given the comfort and familiarity of users with the interface and experience of custodial wallets, we hope to see MFKDF adopted as a primary key management approach for cryptocurrency wallets and thus drive adoption of self-custody solutions.

\bigskip

\noindent \textbf{Limitations}. 
It is expected that the MFKDF-based wallet constructions presented in this paper are only as secure as the underlying factors. For example, if SMS OOBA is used as a factor, but the underlying device is vulnerable to a SIM-swapping attack, then the wallet would be equally vulnerable.
This is equivalent to centralized exchanges, where accounts can be compromised by adversaries knowing the credentials.


While the attestation mechanism of Fig. \ref{fig:system_attestation} serves to increase the cost of a brute-force attack, it does not entirely eliminate its possibility. For example, an attacker who wishes to consume 100~GB of network storage space could create 500,000 wallets, each consuming 200~KB, at a cost of about \$2.5 million USD at current ETH and gas prices. The cost of such an attack could be increased by having a nominal minimum wallet value (e.g., 0.05 ETH), or by using sharded storage of wallet parameters.

Regrettably, we were not able to conduct a full usability study as part of this work, instead focusing on the security and privacy aspects along with a proof-of-concept implementation. We hope to see future work that compares the usability of the proposed wallet design to conventional self-custody wallets.

\section{Future Work}
\label{sec:future}

\subsection{Security}
In addition to the factors used in our implementation (e.g., YubiKey) and the other factors currently supported by MFKDF (e.g., HOTP, TOTP, OOBA), the MFKDF paper \cite{https://doi.org/10.48550/arxiv.2208.05586} also suggests that secure multi-party computation (MPC) could be used to construct additional MFKDF factors, including factors corresponding to geolocation, device identifiers, behavioral authentication, OAuth/OIDC, U2F, and more.
In theory, the policy-based framework of MFKDF would then allow these factors to be combined in arbitrary ways, allowing for expressive policies such as ``require 2 factors if a user is on a familiar device, and 3 factors otherwise.'' When used in combination with existing risk management frameworks \cite{eyal2022cryptocurrency}, this could allow for highly flexible customization of the factors (and combinations thereof) used for authentication while strictly managing and quantifying the implied risk.

\subsection{Privacy}
Next, there are several improvements that can be made to enhance the privacy of the wallet system. Firstly, while the present proposal uses decentralized email verification as a means of providing human-readable identifiers, linking wallets to email addresses may be disadvantageous from a privacy perspective despite the existence of anonymous email services. A suitable alternative may be to use a decentralized namespace such as Ethereum Name Service (ENS) \cite{ens} to provide human-readable but anonymous usernames, though doing so in a way that is not costly to users may require further research.

Additionally, the suggested method uses a ``proof of value'' approach to ensure high availability of legitimate wallets while avoiding denial-of-service attacks that may result from storing an excessive volume of worthless public material. While this strategy is sound in practice, its current implementation may compromise user privacy by causing cryptocurrency wallet addresses to be linkable to the corresponding MFKDF public material. Thus, alternative approaches, such as using Zero Knowledge (ZK) proofs to validate the legitimate holdings of a wallet without revealing the underlying cryptocurrency address, could provide the same security with increased privacy.

Lastly, while moving away from centralized custodial wallets is inherently advantageous from a privacy standpoint, the current system may still reveal the identity or IP address of a wallet owner when public material is requested from the network to perform a ``login'' operation. As such, techniques from the fields of oblivious memory or private information retrieval may be used to obfuscate wallet access and thus enhance user privacy. Private networking protocols such as Tor \cite{dingledine_tor_2004} or Dandelion \cite{fanti_dandelion_2018} can also be used to anonymize network requests, such as when broadcasting a transaction.

\subsection{Scalability}
Lastly, there are further optimizations that could be developed to improve the scalability of the proposed system. While the size of public parameters stored for an individual wallet is usually quite small ($\leq 10$~kb), the storage space required for all users to store all valid wallets could become prohibitive if the system is adopted by millions. Thus, distributed storage techniques, such as sharding, may become necessary at scale. Existing technologies purpose-built for this use case, such as Filecoin \cite{filecoin}, can also be used as the underlying storage solution. Lastly, some factors, such as TOTP, may require $200$~kb or more of storage per user depending on configuration parameters, and the use of compression or other techniques to reduce the amount of data storage may be necessary.
\section{Conclusion}
In this paper, we have presented an initial design for a cryptocurrency wallet based on multi-factor key derivation \cite{https://doi.org/10.48550/arxiv.2208.05586}. Our work is motivated by the observation that existing custodial and non-custodial wallet designs each have significant drawbacks, which we sought to rectify through a secure and user-friendly ``best of both worlds'' approach.

By using MFKDF to derive a wallet key on the fly from standard, unmodified authentication factors (such as passwords and software or hardware-based OTPs), we obviate the need to store private keys at all in any location.
Users can simply ``log in'' to their wallet using their normal authentication factors and re-derive their wallet key as needed, providing the look and feel of a centralized experience with the security of multi-factor authentication. Threshold-based MFKDF allows the wallet key to be recovered even if a subset of the initially established authentication factors are forgotten.
Any public material that requires persistence can be safely stored by peers without requiring a trusted committee, providing fault tolerance, redundancy, and high availability while ensuring that wallets can be accessed from any device.
Thus, the proposed approach succeeds at the stated goals of \S\ref{sec:problem}, inheriting the decentralization and trustlessness of a self-custody solution while providing the portability, resilience, recoverability, and multi-factor authentication (with existing, familiar authentication factors) of a custodial wallet.

The wallet design of this paper is a quintessential example of a setting in which one would never consider using password-based key derivation (e.g., PBKDF2 \cite{rfc2898}) alone. Passwords are known to be a poor solitary authentication factor in most cases \cite{password_reuse, florencio_large_2006}, with the risk of attacks such as credential stuffing \cite{credential_stuffing} being far too high when the consequences include the theft of stored cryptocurrencies. Furthermore, the lack of a secure recovery when using passwords alone means that a forgotten password could entail the total loss of funds, which has unfortunately already occurred in several known instances \cite{popper_lost_2021, radio__this_2021}. Multi-factor key derivation (MFKDF) is the critical improvement that provides significantly stronger security and brute-force attack resistance than password-based key derivation while also natively supporting secure key recovery in case of forgotten factors, allowing for a true custodial-like experience to be achieved in a decentralized way. Thus, we hope the current proposal serves as a turning point in the adoption of self-custody solutions amongst users who presently cling to custodial wallets for usability reasons.


\bibliographystyle{ieeetr}
\bibliography{900-References}

\section*{Acknowledgments}
We appreciate the advice and feedback of Deevashwer Rathee, Xiaoyuan Liu, and Julien Piet. This work was supported in part by the National Science Foundation (NSF), by the National Physical Science Consortium (NPSC), by the Fannie and John Hertz Foundation, and by the Berkeley Center for Responsible, Decentralized Intelligence (RDI). Any opinions, findings, and conclusions or recommendations expressed in this material are those of the authors alone, and do not necessarily reflect the views of the supporting entities.

\section*{Availability}
We invite readers to try the MFKDF wallet demo at \linebreak \url{https://wallet.mfkdf.com}.
The source code for the demo is available at \url{https://github.com/multifactor/mfkdf-wallet-demo}.

\end{document}